\documentclass[aps,twocolumn,superscriptaddress,amsfont,graphicx,nofootinbib,preprintnumbers]{revtex4-1}%
\UseRawInputEncoding
\usepackage{amsmath, amsthm, amssymb}
\usepackage{graphicx}
\usepackage{subfigure}
\usepackage{wrapfig}
\usepackage{color}
\usepackage{color,graphicx,epsfig}
\usepackage{ifpdf}
\usepackage{bm}
\usepackage[english]{babel}
\usepackage{braket}
\usepackage{hyperref}
\usepackage{enumerate}
\usepackage{url}
\usepackage{multirow}
\usepackage{tablefootnote} 
\usepackage{cases} 
\usepackage{siunitx}

\usepackage{slashed}

\usepackage{changes}

\begin{document}
\title{Can Orbital Decay of Accreting Binary Pulsars Probe Dark Matter?}

\author{
	Arvind Kumar Mishra 
	${\href{https://orcid.org/0000-0001-8158-6602}{\includegraphics[height=0.15in,width=0.15in]{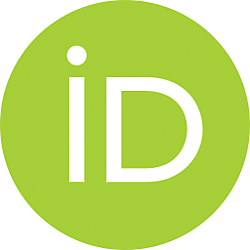}}}$ 
}
\email{arvindm215@gmail.com}
\affiliation{Department of Physics and Institute of Theoretical Physics, Nanjing Normal University, Nanjing, 210023, China}
\affiliation{Nanjing Key Laboratory of Particle Physics and Astrophysics, Nanjing, 210023, China}

\date{\today}

\begin{abstract}
The merger of binary pulsars in dark matter (DM)-rich environments can result in DM particle accretion, leading to an increase in the individual pulsar masses. In this work, we investigate the effects of DM accretion on the change in orbital period rate of binary pulsars. Our analysis reveals that while DM accretion increases the system's mass, it may also modify the orbital evolution by enhancing the orbital decay rate. By comparing our results with existing binary pulsar data near Earth's location, we report that the current DM accretion rate is insufficient to place meaningful constraints on DM particle properties. However, we demonstrate that future observations of pulsar mergers in the high DM-density environment of the galactic center could offer a unique opportunity to probe DM microphysics through this mechanism.
\end{abstract}
		

		\maketitle

\section{Introduction}
The existence of dark matter (DM) has been firmly
established through a wide range of observations, spanning from small to large cosmological scales~\cite{Bertone:2004pz}. However, while all current evidence for DM is gravitational in nature, its microphysical properties remain an enigma. Despite extensive efforts, ongoing DM detection experiments have not yet observed dark matter particles but have instead placed stringent constraints on their microphysics \cite{Klasen:2015uma,Liu:2017drf,PerezdelosHeros:2020qyt}.

The high densities and extreme gravitational fields of the
compact objects (white dwarfs, neutron stars, and black holes) make them ideal natural laboratories for
studying DM particles (see a recent review \cite{Bramante:2023djs}).
When located in DM-dominated environments, these compact
objects can efficiently capture DM particles, potentially
altering the internal structure of compact objects. The captured DM particles may thermalize within the star and lead to additional heating. Consequently, observations of
old, isolated, and anomalously hot neutron stars could
provide valuable constraints on various DM models \cite{Reisenegger:2006ky,Kouvaris:2007ay,Raj:2017wrv,Bell:2018pkk,Yanagi:2019vrr,Hamaguchi:2019oev,Bell:2020lmm,Garani:2020wge,Joglekar:2020liw,Fujiwara:2022uiq,Baryakhtar:2017dbj,Ema:2024wqr,Linden:2024uph}. In cases of high DM accretion, where the accumulated DM exceeds the Chandrasekhar limit within the thermalisation radius, the DM core may collapse to form a black hole. This newly formed black hole could then accrete
neutron star material, ultimately leading to the destruction of the host star \cite{Goldman:1989nd, 1990PhLB..238..337G,Kouvaris:2010vv, deLavallaz:2010wp, McDermott:2011jp,Bramante:2014zca,Garani:2018kkd,Bramante:2013nma,Bramante:2013hn, Bell:2013xk, Guver:2012ba, Garani:2018kkd, Kouvaris:2018wnh,Lin:2020zmm,Garani:2021gvc,Singh:2022wvw}. Therefore, observations of
old neutron stars situated within DM-dominated regions offer
a promising avenue for probing the microphysical properties of dark matter. Moreover, the strong magnetic fields
surrounding compact objects provide complementary opportunities to investigate novel DM candidates, including
axions \cite{Hook:2018iia,Song:2024rru,Carenza:2024ehj,Caputo:2024oqc,Roy:2025mqw}, dark photons \cite{Marocco:2021dku,Yan:2023kdg,Sulai:2023zqw}, and millicharged DM
particles \cite{Gong:2025xsd,Arza:2025cou}. For comprehensive discussions on DM interactions with compact objects, we refer
Refs. ~\cite{Bertone:2007ae,Kouvaris:2010jy,2011PhRvD..84j7301L,Gresham:2018rqo,Deliyergiyev:2019vti,Acevedo:2019gre,Dasgupta:2020mqg,Dasgupta:2020dik,Kain:2021hpk,Bramante:2022pmn,Nguyen:2022zwb,Coffey:2022eav,Ray:2023auh,Bhattacharya:2023stq,Niu:2024nws} and references therein.

Furthermore, binary mergers have been utilized to probe dark matter microphysics. Current studies primarily focus on black hole binaries are used to probe the DM properties via superradiance effects  \cite{Siemonsen:2022ivj,Chen:2025jch,Alonso-Alvarez:2024gdz,Ding:2025nxe,Li:2025qyu}.
Additionally, neutron star-black hole binary systems have emerged as promising probes of DM microphysics \cite{Akil:2023kym}. Recent work has also demonstrated the growing potential of binary mergers (including white dwarf-white dwarf, white dwarf-neutron star, and NS-NS) for such investigations via DM dynamical friction \cite{Gomez:2019mtl}.

In this work, we investigate the properties of dark matter through observations of merging pulsars embedded in a DM-dominated environment. When a binary pulsar merges in such an environment, it can accrete DM particles and therefore increase the individual pulsar masses. This accretion process may consequently alter the orbital period evolution of the binary system. We calculate the DM mass capture rate on neutron stars through multiscatter interactions, considering velocity-dependent DM-baryon elastic scattering. Our analysis reveals that a positive velocity dependence enhances the capture rate, particularly for low DM masses and large scattering cross-sections, thereby influencing the orbital period evolution. Using observational data from pulsar binaries, we assess the potential to probe DM properties. Our results indicate that while DM accretion increases the binary masses, its contribution is too negligible to significantly modify the orbital decay rate. As a result, current data do not provide meaningful constraints on DM microphysics. We note that existing pulsar binary observations are predominantly located far from the galactic center, where DM densities are lower. However, we emphasize that future observations of binary pulsar mergers near the galactic center, where DM densities are highest, could offer a promising avenue to explore DM interactions using this method.

The arrangement of this work is as follows: In Section \ref{sec:orbit}, we derive the expression of the orbital decay rate due to gravitational waves and dark matter accretion. In Section \ref{sec:capture}, we discuss the DM capture rate using velocity dependent DM-Nucleon elastic cross-section via multiscatter capture formalism. Further, in Section \ref{Sec:results}, we present and discuss our results. Finally, we conclude our findings in Section \ref{Sec:conc}.
\section{Evolution of Orbital Period due to GW and DM accretion  }
\label{sec:orbit}
We assume that a binary pulsar system with masses $M_1$ and $M_2$ is orbiting in a Kepler orbit. The binary system is described with total energy, $E = -G\mu M/2a$, and orbital period, $P=(4\pi^{2}a^{3}/GM)^{1/2}$, in which $M=M_{1}+M_{2}$ is total mass, $\mu =M_{1}M_{2}/(M_{1} +M_{2})$ is reduced mass, and $a$ represents the semi-major axis. Using this, orbital period of binary, $P$ can be computed as
\begin{equation}
P = \frac{2\pi GM \mu^{\frac{3}{2}}}{(-2E)^\frac{3}{2}}\, .
\label{eq:period}
\end{equation}
From the above equation, we see that the orbital period depends on the total energy and masses of the orbiting pulsars. Therefore, the ratio between the rate of change of the orbital period with respect to time ($\dot{P}$) and the orbital period ($P$), $\dot{P}/P$, can be estimated using the above equation as 
\begin{equation}
\frac{\dot{P}}{P} = -\frac{3}{2}\frac{\dot{E}}{E} + \left(1+\frac{M_{2}}{2M}\right)\frac{\dot{M_{1}}}{M_{1}} +\left(1+\frac{M_{1}}{2M}\right)\frac{\dot{M_{2}}}{M_{2}}~,
\label{eq:orbit}
\end{equation}
where $\dot{E}$ is the average energy decay rate due to GW emission, and $\dot{M_{1}}$ and $\dot{M_{2}}$ correspond to the accreted mass by the binaries.
Furthermore, we consider that both neutron stars in the binary are in the DM-rich region and accrete surrounding DM particles. Then the DM capture rate for an individual pulsar is given by
\begin{equation}
\dot{M_{1}}=  m_{\chi} C_{M_1},  \dot{M_{2}}=m_{\chi} C_{M_2}~,
\end{equation}
where $m_{\chi}$ is DM mass, and $C_{M_1}$ and $C_{M_2}$ are DM capture rates for the first and second pulsars, respectively. Using the above, the total decay rate will be given by
 \begin{equation}
\frac{\dot{P}}{P} = -\frac{3}{2}\frac{\dot{E}}{E} + \left(1+\frac{M_{2}}{2 M}\right)\frac{m_{\chi}C_{M_1}}{M_{1}} +\left(1+\frac{M_{1}}{2 M}\right)\frac{m_{\chi}C_{M_2}}{M_{2}}~.
\label{eq:orbitfinal}
\end{equation}
For equal binary pulsar masses, i.e., $M_1=M_2=M$, and $C_{M_1}=C_{M_2}=C_{M}$, the above equation reduces as 
\begin{equation}
\frac{\dot{P}}{P} = -\frac{3}{2}\frac{\dot{E}}{E} + \frac{5}{2}\frac{m_{\chi}C_{M}}{M}~.
\label{eq:orbitfinaleq}
\end{equation}
Further, the average rate of energy emission (due to gravitational wave emission) in the post-Newtonian expansion to order zero is given by
\cite{Peters:1963ux,Peters:1964zz,Blanchet:1989cri,Maggiore:2007ulw}
\begin{equation}
\dot{E} = -\frac{32G^4}{5c^5}\frac{\mu^{2} M^3}{a^5(1-e^2)^\frac{7}{2}}\left(1+\frac{73}{24}e^2+\frac{37}{96}e^4+...\right), \label{eq:dotE}
\end{equation}
where $G$, $c$, and $e$ represent Newton's gravitational constant, the speed of light, and the orbital ellipticity, respectively. Furthermore, the total energy can be expressed in terms of the orbital period as $E=-\frac{\mu}{2^{1/3}}(\frac{GM \pi}{P})^{2/3}$. 

From Eq.~(\ref{eq:dotE}), we find that gravitational wave emission decreases the orbital decay rate as it carries energy away from the binary system. This effect, first observed in the Hulse-Taylor binary pulsar PSR B1913+16~\cite{Hulse:1974eb}, provided the first indirect evidence for GWs. Conversely, dark matter accretion increases the orbital decay rate by transferring angular momentum to the binary. These two competing processes can reach equilibrium when their effects cancel each other, resulting in a stationary orbital period. This equilibrium scenario may occur in systems with high accretion rates (such as in accreting binaries) and has been proposed as a source of continuous GWs through torque balance~\cite{Bildsten:1998ey}.

Furthermore, as the energy emission rate via GW emission is known, to calculate the theoretical decay rate, we need to estimate the DM capture rate via a pulsar. 

\section{\label{sec:accretion} Capture rate due to DM-nucleon interactions}
\label{sec:capture}
We consider that the binary pulsars are surrounded by dark matter particles. Here, we consider asymmetric dark matter and neglect any possible self-interaction between the DM particles. Further in this work, we focus on the larger DM masses $m_{\chi}>$ GeV; therefore, we safely neglect the evaporation effect. The DM capture rate $C_{\chi\mathrm{N}}$ has been explored both in single scattering  Refs.~\cite{1985ApJ...296..679P,1987ApJ...321..571G,Bell:2020obw,Robles:2022llu,Bell:2023ysh}, and multiscattering scatter process \cite{Bramante:2017xlb,Steigerwald:2022pjo,Leane:2023woh}. For a detailed discussion, see a recent review  \cite{Bramante:2023djs}.

Here we focus on DM capture on pulsar via multiscatter capture process, which has been explored in detail in Refs. \cite{Bramante:2017xlb}. In this scenario, the capture rate due to $\mathrm{N}$ number of scattering, $C_{\mathrm{N}}$ is given by \cite{Bramante:2017xlb} 
\begin{equation}
C_{\mathrm{N}}=  \pi R^{2}_{\mathrm{NS}}\int^{\infty}_{0} \frac{f(u)}{u}~ w^{2} g_{\mathrm{N}}(w)~du~~,
\label{eq:Cnx}
\end{equation}
where $f(u)$ is the DM velocity distribution function, and $w= \sqrt{u^{2}+v^{2}_{\mathrm{esc}}}$ in which $u$ is DM velocity at very large distance and $v_{\mathrm{esc}}$ is escape velocity. 
$g_{\mathrm{N}}(w)$ represents the probability of reducing DM particle velocity lower than the escape velocity after $N$ collisions. Here, we approximate, $g_{\mathrm{N}}(w)=\Theta \left[v_{\mathrm{esc}}(1-\langle z_{i}\rangle \beta_{+}/2)^{-N/2}-w\right]$, where, $\beta_{+}=4m_{\mathrm{N}} m_{\chi}/(m_{\mathrm{N}}+m_{\chi})^{2}$ in which, $m_{\chi}$ and $m_{N}$, represents the mass of the DM and nucleons, respectively \cite{Bramante:2017xlb}. 
 \begin{figure*}[]
    \centering
    \includegraphics[width=0.9\columnwidth]{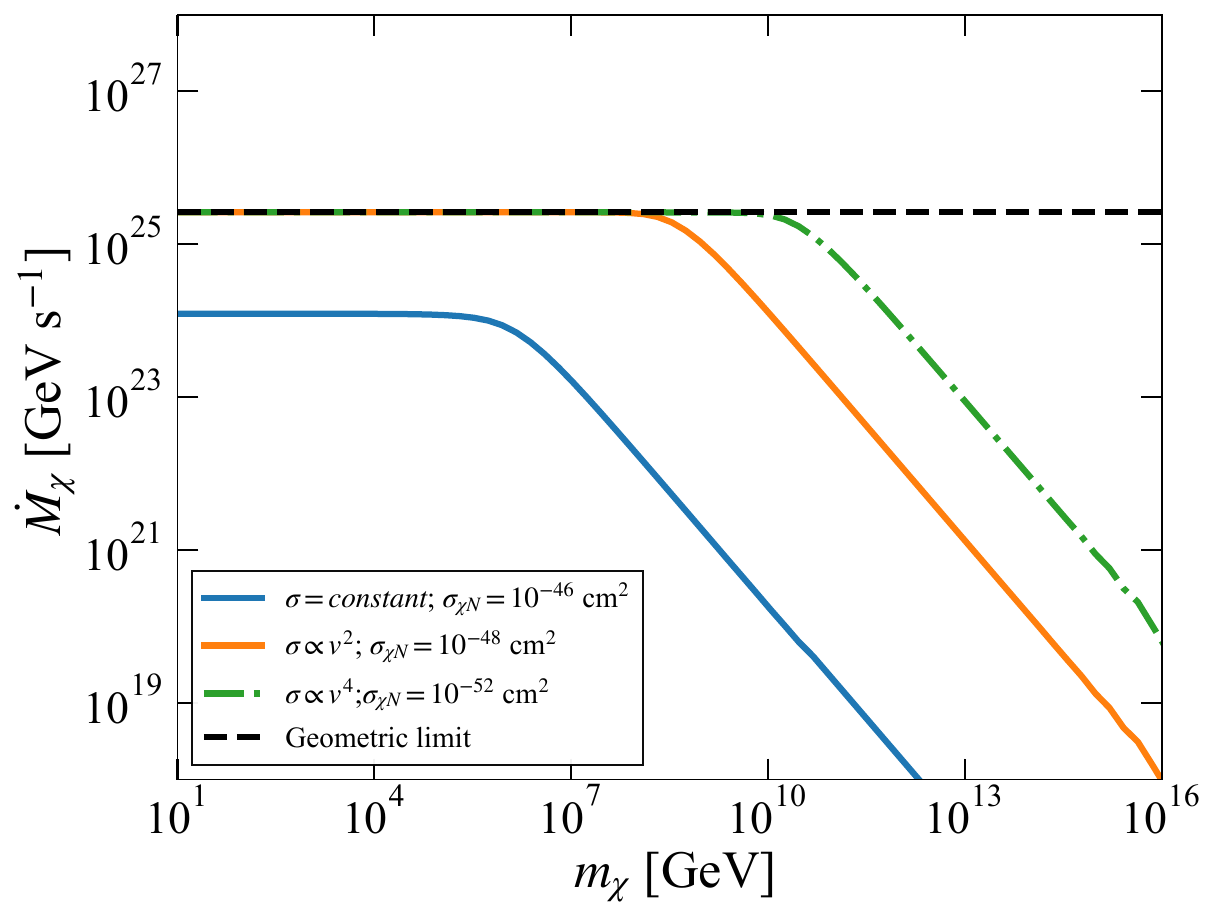}    \includegraphics[width=0.9\columnwidth]{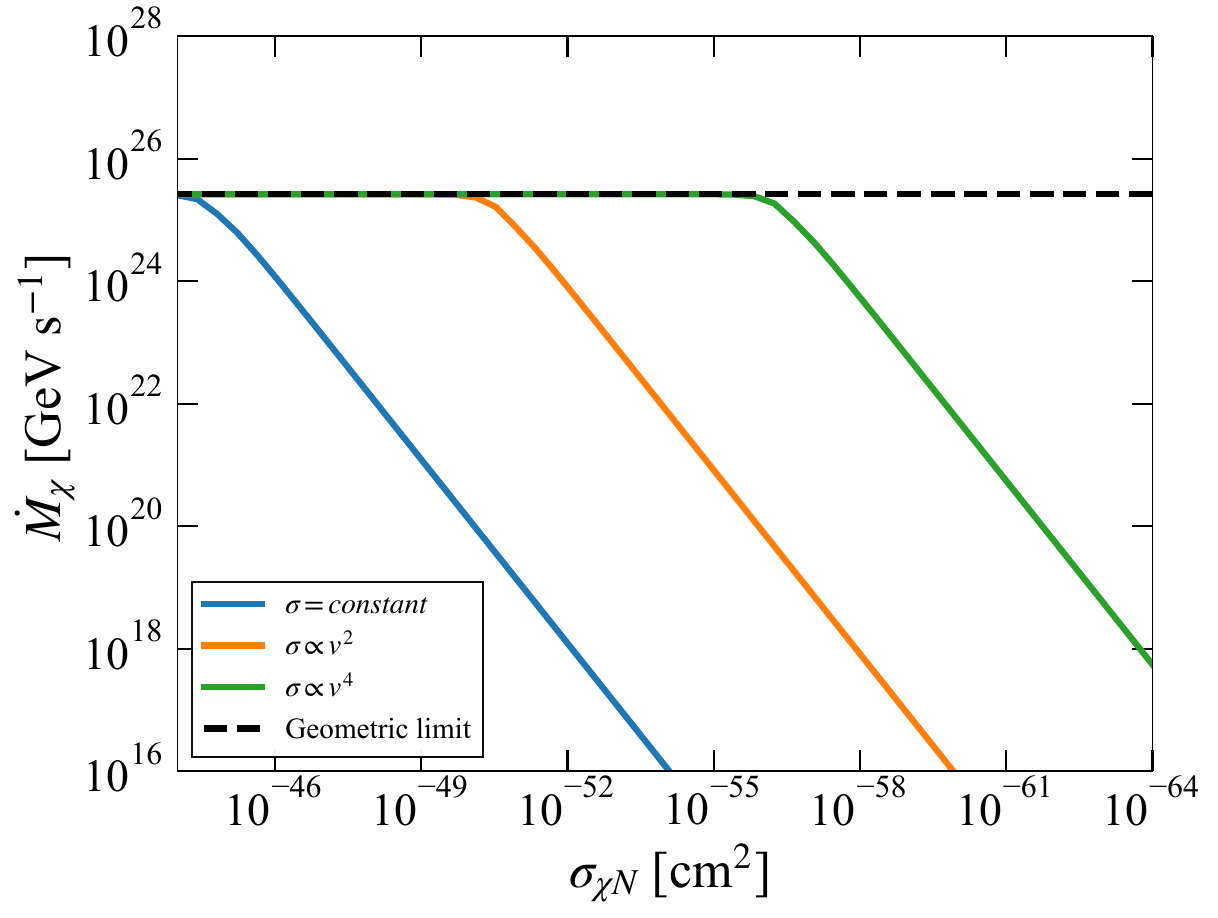} 
    \caption{ The DM mass capture rate on the binary pulsar due to DM accretion. \textit{Left panel} and \textit{Right panel} represent capture rate as a function of DM mass, and cross-section, respectively. The geometric limit (dotted) represents the maximum possible rate. }
    \label{fig:capture}
\end{figure*}
 Further, $p_{\mathrm{N}}(\tau)$ corresponds to the probability of DM particles participating in $N$ collisions with an optical depth, $\tau$.  $p_{\mathrm{N}}(\tau)$ can be defined via Poisson distribution as \cite{Bramante:2017xlb}
\begin{equation}
 p_{\mathrm{N}}(\tau)=2\int^{1}_{0} y e^{-y\tau}  \frac{(y\tau)^{\mathrm{N}}}{\mathrm{N}!} dy~~,
 \label{eq:probability} 
\end{equation}
where $y$ is related to the incidence angle of DM particles and $\tau=3\sigma/(2\sigma_{\mathrm{geo}})$, where $\sigma$ is the DM-nucleon cross-section and $\sigma_{\mathrm{geo}}$ is the geometrical cross-section. For a pulsar, the geometrical cross-section is given by  $\sigma_{\mathrm{geo}}=\pi R^{2}_{\mathrm{NS}}/N_{\mathrm{B}}$ where, $R_{\mathrm{NS}}$ is the radius of NS, and $N_{\mathrm{B}}$ is total number of the baryons, respectively. Further, DM particle velocity distribution function is assumed to be Maxwellian, and defined as (in the DM halo rest frame) \cite{1985ApJ...296..679P} 
\begin{equation}
 f(u) = \left(\frac{3}{2} \right)^{\frac{3}{2}} \frac{4}{\sqrt{\pi}} \frac{\rho_{\chi}}{m_{\chi}} \frac{u^{2}}{u^{3}_{0}}e^{-\frac{3}{2}\frac{u^{2}}{u^{2}_{0}}}~,
 \label{eq:dist}
 \end{equation}
 where $\rho_{\chi}$ is the DM energy density and $ u_0$ is average velocity of the DM particles. For calculation purposes, we consider $ u_0 =270$km$/$s. Therefore, the total capture rate after the $N$ collision is obtained as
\begin{equation}
C_{\mathrm{total}}=\sum^{\infty}_{\mathrm{N}=1} C_{\mathrm{N}}~.
\end{equation}

To calculate the capture rate, we need to provide the model for the DM-nucleon scattering. Here, for simplicity, we explore the constant and velocity-dependent cross-section (VDCS). We parameterize VDCS as a power law form, given by \cite{Vincent:2015gqa,Lu:2024kiz}
\begin{equation}
\sigma(v_{\mathrm{rel}})=\sigma_{\chi \mathrm{N}}\left(\frac{v_{\mathrm{rel}}}{u_{\mathrm{ref}}}\right)^{2\alpha}\simeq\sigma_{\chi \mathrm{N}}\left(\frac{w}{u_{\mathrm{ref}}}\right)^{2\alpha},
\label{eq:vdcs}
\end{equation}
where, $v_{\mathrm{rel}}$ is the relative velocity between DM-nucleon, and $u_{\mathrm{ref}}$ is reference velocity. In this parametrization, $\alpha$ depends on the DM models, and for simplicity, we consider $ \alpha=1,2$, which will enhance the DM capture rate as discussed in previous works \footnote{In this work, we are not considering negative velocity dependence and the momentum dependence cross-section case. This is because the capture rate for both cases is generally smaller than the positive velocity dependence cases \cite{Lu:2024kiz,Liu:2024qbe}. We aim to explore the maximum capture rate, so leave this possibility.} \cite{Lu:2024kiz,Liu:2024qbe}. In the particle physics model, $\alpha=1$ (p-wave) and $\alpha=2$ (d-wave) originate whenever the relative angular momentum of initial state particles has 1 or 2 units, see Refs. \cite{Vincent:2015gqa,Kumar:2013iva}. For our estimation, we assume $ u_{\mathrm{ref}} =270$km$/$s.

 Furthermore, using the general expression of probability, the estimation of the capture rate becomes tedious and generally leads to unstable results. Thus, to speed up our calculation, we follow the approach as discussed in Ref. \cite{Leane:2023woh}. We restrict sum limit $N=(10, e \tau)$ \cite{Ilie:2020vec}. In addition, for extreme cases of optical depth, we adopt a simplified expression of probability as  
 \cite{Bramante:2023djs}
\begin{subnumcases}
{P_{\mathrm{N}} (\tau)\approx} 
\frac{2\tau^{\mathrm{N}}}{N!(N+2)}+\mathrm{O}(\tau^{\mathrm{N}+1}), \quad if~~ \tau\ll 1\label{eq:CtotSSHM}\\
\frac{2}{\tau^{2}}\left(N+1\right) \Theta \left(\tau-N\right), \quad if~~ \tau\gg 1 \label{eq:CtotSSLM}
\end{subnumcases}
Using the capture rate formalism described above, we compute the accreted dark matter particles employing the \textit{Asteria} package \cite{Leane:2023woh}. This numerical tool calculates DM capture rates in compact objects, originally designed for constant cross-section interactions. We use it for the neutron star and modify it for the VDCS case. To modify the code for the velocity cross section, we follow a conservative approach and use a simplified approximation for the cross section. We approximated the cross-section via  $\sigma(v_{\mathrm{rel}})\approx \sigma_{\chi \mathrm{N}}\left(w_{0}/u_{\mathrm{ref}}\right)^{2\alpha}$, where, $w_{0}= \sqrt{u^{2}_{0}+v^{2}_{\mathrm{esc}}}$. This simplified approximation works well and provides stable results for VDCS also.

\section{Results}
\label{Sec:results}
After being equipped with the necessary ingredients, we now show our final results. 

\begin{figure*}
    \centering
    \includegraphics[width=0.9\columnwidth]{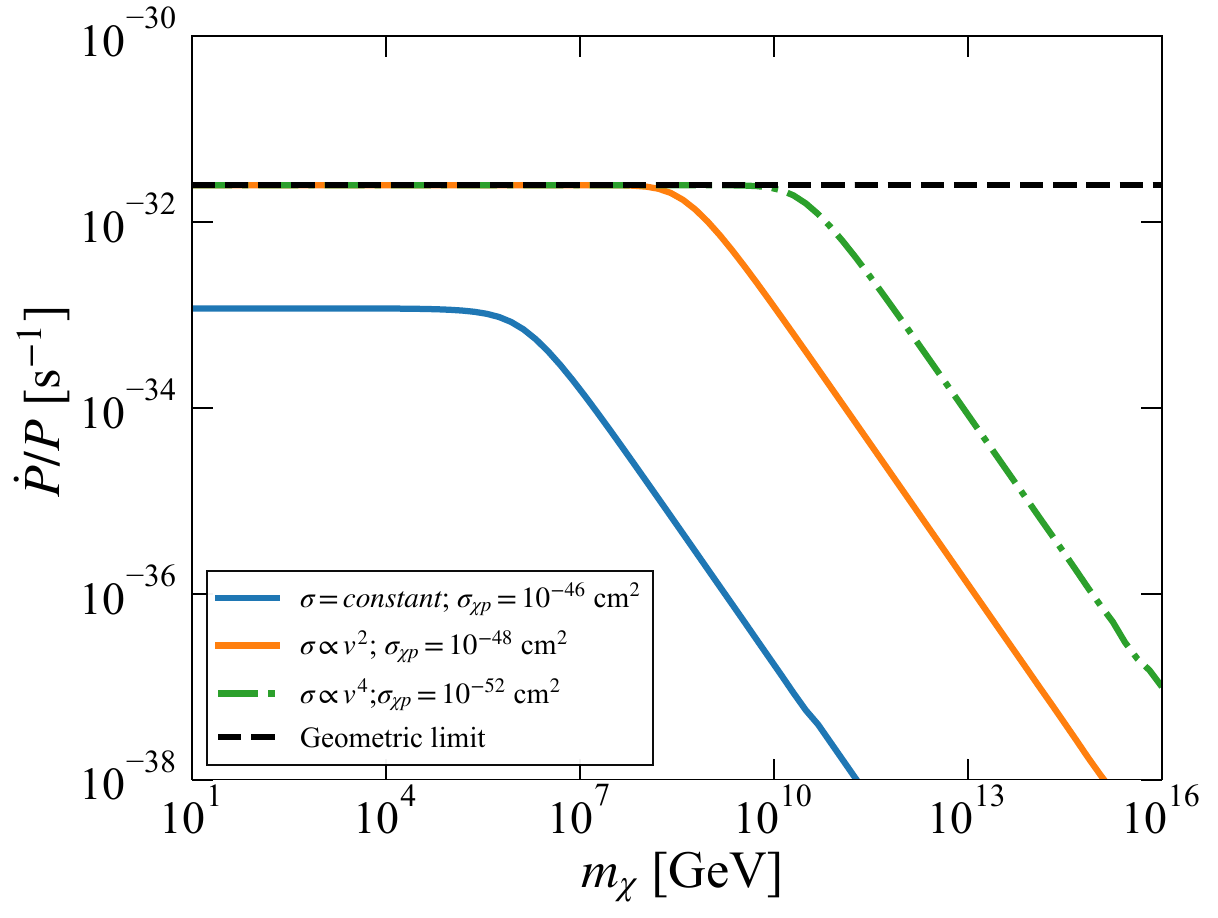}    \includegraphics[width=0.9\columnwidth]{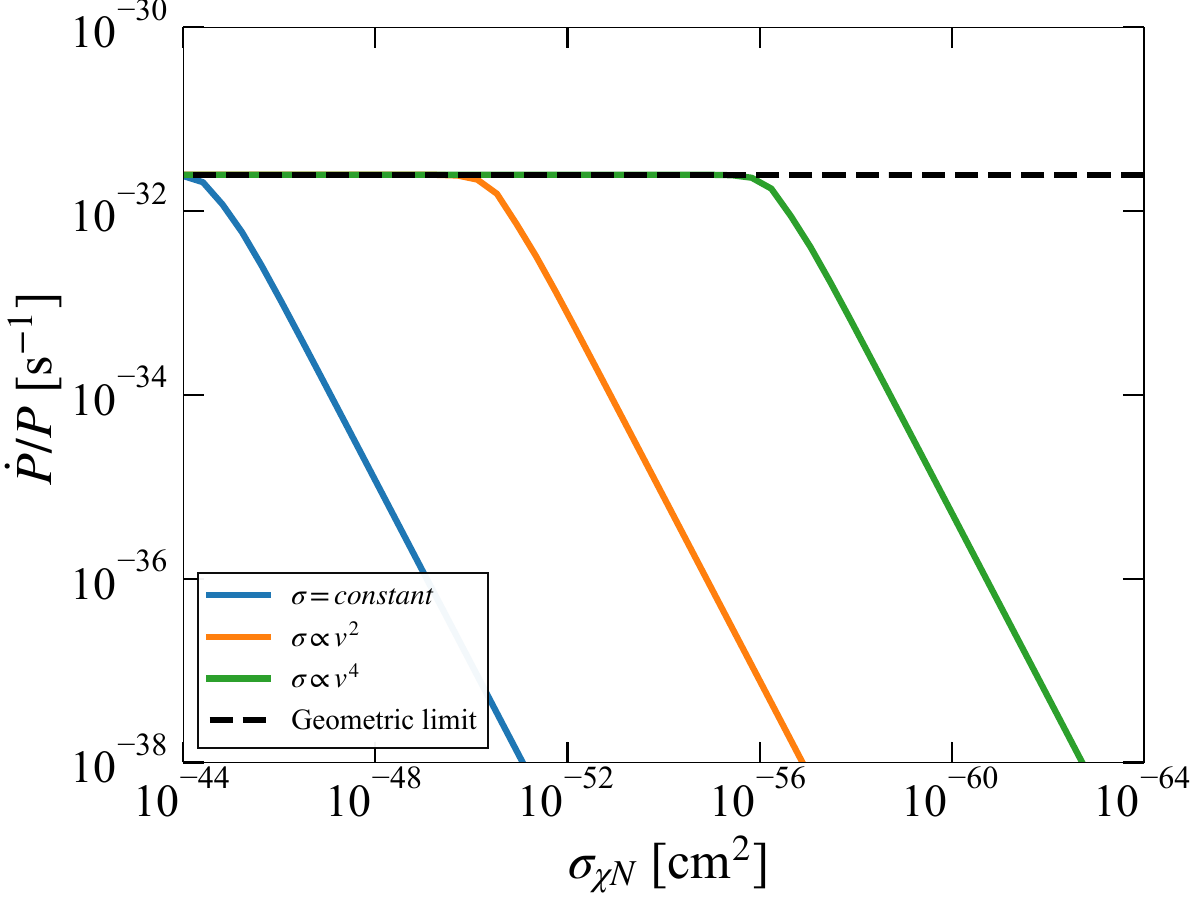}
   \caption{ The normalized decay rate of orbital period, $\dot{P}/P$ of binary pulsar due to DM accretion. \textit{Left panel} and \textit{Right panel} represent capture rate as a function of DM mass, and cross-section, respectively. The geometric limit represents the maximum possible capture rate. } 
    \label{fig:period}
\end{figure*}
\begin{table*}
\begin{center}
\begin{tabular}{|c | c | c | c | c | c | c | c | c | }
\hline  Pulsar & $P$ (days) & $\dot{P}$ & Eccentricity & $M_{1}(M_{\odot})$ & $M_{2}(M_{\odot})$ & Distance (kpc) & Reference\\ 
\hline
 B1534+12  & $0.420737298881(2)$ & $-0.19244\times10^{-12}$ & $0.27367740(4)$ & $1.3330(2)$  & $1.3455(2)$   & 1.051& \cite{Fonseca:2014qla}\\ 
\hline
  B1913+16 & $0.322997448918(3)$ & $-2.398(4)\times10^{-12}$ & $0.6171340(4)$ & $1.390(1)$ & $1.438(1)$ & 0.34&\cite{Weisberg:2016jye}\\ \hline
 J0737-3039 & $0.10225156248(5)$ & $-1.252(17)\times10^{-12}$ & $0.0877775(9)$ & $1.2489(7)$ & $1.3381(7)$ & 0.5 &\cite{Kramer:2006nb} \\ \hline 
  J1757-1854 & $0.18353783587(5)$ & $-5.3(2)\times10^{-12}$ & $0.6058142(10)$ & $1.3946(9)$ & $1.3384(9)$ &0.6&\cite{Cameron:2017ody}  \\ \hline
\end{tabular}
\caption{The measured orbital parameters for the binary pulsar systems. The distances are given from the Earth.}
\label{tab:data}
\end{center}
\end{table*} 
\begin{table*}
\begin{center}
\begin{tabular}{|c | c | c | c | c | c | c | c | c | }
\hline  Pulsar & $\dot{P}/P|_{obs}$ (sec) & $-2\dot{E}/3E|_{GW}$(sec)& $\dot{P}/P|_{DM}$ (sec)\\ \hline
 B1534+12  & $-5.294\times10^{-18}$ & $-5.293\times10^{-18}$&$1.835\times10^{-33}$  \\ \hline
  B1913+16 & $-8.593\times10^{-17}$ & $-8.606\times10^{-17}$  & $1.938\times10^{-33}$  \\ 
\hline
 J0737-3039 & $-1.417\times10^{-16}$ & $-1.412\times10^{-16}$ & $1.771\times10^{-33}$  \\ \hline
  J1757-1854 & $-3.342\times10^{-16}$  & $-3.325\times10^{-16}$ & $1.872\times10^{-33}$   \\ \hline
\end{tabular}
\caption{The contribution of the DM accretion, GW, and observed from the pulsar binary $\dot{P}/P|_{obs}$. The theoretical strength of the DM contribution has been quoted using our method. }
\label{tab:result}
\end{center}
\end{table*} 
\subsection{DM mass accretion rate }
Fig. \ref{fig:capture} shows the mass of accreted  DM accretion on a pulsar for constant and velocity-dependent cross-section. Here we consider pulsar mass, $M=1.4 M_{\odot}$, and $R_{\mathrm{NS}}=10$ km. \textit{Left panel} refers to the capture rate as a function of the DM mass, keeping the cross-section fixed. Here, green dot-dashed, orange solid, and blue color lines correspond to the constant, $\sigma\propto v_{\mathrm{rel}}^{2}$ and $\sigma\propto v_{\mathrm{rel}}^{4}$ models, respectively. The black dashed line corresponds to the geometric limit, which represents the maximum possible DM capture rate. We observe that for lower DM masses, the capture rate is constant and decreases for higher DM masses, which is expected. For positive velocity dependence, although the cross-section is small (i.e., $\sigma_{\chi N}=10^{-48}$cm$^{2}$ for $\alpha=1$, and $\sigma_{\chi N}=10^{-48}$cm$^{2}$ for $\alpha=2$ model), but mass capture rate is higher than the constant cross-section case (for which $\sigma_{\chi N}=10^{-46}$cm$^{2}$). For larger DM masses, the capture rate is largest for quartic velocity dependence, $\sigma\propto v_{\mathrm{rel}}^{4}$. 

\textit{Right panel} of Fig. \ref{fig:capture}, shows the mass capture rate as a function of the cross-section, keeping DM mass fixed  ($m_{\chi}=10^{5}$ GeV). Here, the color lines correspond to the same models as discussed before. We see that for lower cross sections, the capture rate saturates to a geometrical rate and is reduced for higher DM masses. For a larger cross-section case, the capture rate is higher for positive velocity dependence in comparison to the constant case as discussed before.

\subsection{Effect of DM accretion on $\dot{P}/P$ }
Now, we show the effect of the DM accretion on $\dot{P}/P$ of binary pulsar as a function of the DM mass in Fig. \ref{fig:period}. Here we have shown only the impact of DM contribution (sum of second and term in r.h.s of Eq. (\ref{eq:orbit})) on $\dot{P}/P$. \textit{Left panel} represents $\dot{P}/P$ as a function of the DM mass for a fixed cross-section. Here, blue, orange, and green dot-dashed color lines correspond to constant, $\sigma\propto v_{\mathrm{rel}}^{2}$ and $\sigma\propto v_{\mathrm{rel}}^{4}$ respectively. On the low DM masses, $\dot{P}/P$ is constant and decreases for higher DM masses.  This can be understood from Eq. (\ref{eq:orbit}), which suggests $\dot{P}/P \propto \dot{M}$.  As accreted mass remains a constant on low DM masses, and decreases on comparably higher masses, $\dot{P}/P$ will also follow the same behavior. Also, on the larger DM masses, the accretion rate is higher for $\sigma\propto v_{\mathrm{rel}}^{4}$ case, and therefore on those masses it will influence $\dot{P}/P$ more strongly. Further, \textit{Right panel} of Fig. \ref{fig:period} represents the $\dot{P}/P$ as a function of the cross section keeping the DM mass fixed ($m_{\chi}=10^{5}$ GeV). As $\dot{P}/P \propto \dot{M}$, therefore the behavior of  $\dot{P}/P$ will depend on mass capture rate as discussed before, see \textit{Right panel} of Fig. \ref{fig:period}. We also point out that the maximum change for $\dot{P}/P$ happens (due to DM accretion) for the geometric limit.
\subsection{Can a decay of the binary pulsar period constrain DM microphysics?}
Now we explore the strength of the change in the orbital rate due to DM accretion using the existing decay rate of the binary pulsars. For this purpose, we use the pulsar merger data from Table \ref{tab:data}.
In this table, we summarize the data from various neutron star mergers, including the rate of change of orbital period, orbital period, masses of the binary pulsar, and their distances from Earth. As all the pulsars are quite near to the Earth, we assume the DM density, $\rho_{\chi}=0.4$GeVcm$^{-3}$.  

Then we compare the $\dot{P}/P$ obtained from the observational data and our theoretical models. In Table \ref{tab:result}, we summarize our theoretical findings along with the observed value, $\dot{P}/P|_{obs}$ from pulsar binaries. Here $-2\dot{E}/3E|_{GW}$, and $\dot{P}/P|_{DM}$ correspond to energy emission due to GW emission and DM contribution in $\dot{P}/P$ (see Eq. (\ref{eq:orbit})). We specifically focus on the $\sigma\propto v_{\mathrm{rel}}^{4}$ DM models for which the mass accretion rate is high and can contribute maximum. Further in the calculation, we fixed $m_{\chi}=10^{5}$ GeV, and $\sigma_{\chi N}=10^{-46}$ cm$^{2}$. We see that for all binary pulsars, the majority of the energy released is carried by the GW radiation, which therefore explains the observed decay period of the pulsar. In this region, the accretion of DM is low and therefore becomes insignificant to modify the orbital period evolution. In fact, for optimal cases, the contribution is at least 15 orders of magnitude smaller than the GW case. Thus, in light of the above, it seems that the current observed binary pulsar can not probe the DM microphysics.

Nevertheless, we also observed that most of the observed binary pulsars are situated quite far from the galactic center, where the DM density is small. Further, as the mass capture rate depends on the DM density $C_{N}\propto \rho_{\chi}$ (see, Eq. (\ref{eq:Cnx}), and Eq. (\ref{eq:dist})), there will be a significant enhancement in the mass capture rate near the galactic center where the DM density is very high. For example, if pulsar merging happens in the DM spike (for which peak density in spike region $\rho = 10^{18}$ GeVcm$^{-3}$ for the non-annihilating DM particles and also lower density for annihilating DM particles \cite{Herrera:2023nww,Mishra:2025juk}), it will enhance the mass, capture rate by the same orders of magnitude (as near Earth $\rho_{\chi}=0.4$GeVcm$^{-3}$). In such a scenario, DM accretion will influence the rate of orbital period, and therefore, in that situation, the observation can constrain the DM properties.  Therefore, future observations of binary pulsar mergers near the galactic center may open a novel opportunity to explore the DM microphysics and galactic environment. 
\section{Conclusion}
\label{Sec:conc}
The binary pulsar merging inside the dark matter-rich environment can accrete the DM particles during its merger and enhance the individual pulsar masses. If the DM mass accretion rate is sufficiently high, it may modify the orbital period rate, and therefore, the observation of the orbital decay rate of the pulsar may be able to shed light on DM microphysics. Motivated by this, in this work, we explore the impact of DM accretion on the orbital decay of the binary pulsar. 

Assuming velocity-dependent DM-baryon elastic scattering, we estimated the DM capture rate on the pulsar via multicapture scattering. The positive velocity dependence enhanced the capture rate for low DM mass and large scattering cross sections, which modified the orbital period decay rate. Further, after using the existing pulsar binary data, we aimed to explore the DM properties. We found that although DM accretion leads to an enhancement of the binary pulsar masses, their contribution is very small (at least 15 orders of magnitude smaller than the GW contribution). Therefore, we reported that the existing binary pulsar observations appear to be unable to probe DM properties.

We further note that existing pulsar binary observations are limited to regions far from the galactic center (near Earth's vicinity), where the ambient DM density is relatively low. A binary pulsar system near the galactic center, where DM densities are significantly higher, would experience enhanced DM capture rates, potentially enabling investigations of DM microphysics. However, current surveys have failed to detect such systems; known as the ``missing pulsar problem" \cite{Fuller:2014rza,Caiozzo:2024flz,Ferrer:2024xwu}. This absence could arise from either (i) physical mechanisms that suppress pulsar formation or survival in the galactic center environment \cite{Fuller:2014rza,Caiozzo:2024flz,Ferrer:2024xwu}, or (ii) observational challenges in detecting faint pulses from this region. While this issue remains unresolved, future advances in observational sensitivity may enable the detection of pulsar mergers near the galactic center, opening new avenues to probe DM properties. Such observations could also constrain models of DM spikes predicted to form in these high-density regions \cite{Quinlan:1994ed,Gondolo:1999ef,Ullio:2001fb}.

Furthermore, there are certain limitations in our estimation. In our work, DM capture rate calculation has been done under certain approximations, such as a non-rotating neutron star, a constant geometrical cross section, and the DM rest frame. We emphasize that in the pulsar frame and a low rotation period (high frequency), the accretion rate is further reduced. When accounting for all the factors together, the capture rate will be modified. These aspects need to be explored further, and we leave it for future work.

\textbf{Note added:} While finalizing this work, we became aware of a similar study exploring DM interactions using binary pulsar mergers \cite{Lucero:2024cfn}. Their work assumes a single-scatter capture scenario and uses existing orbital decay rate data to constrain DM-baryon scattering. In contrast, our analysis does not constrain DM microphysics because, in a multiscatter capture process, the capture rate remains small even for a large DM-baryon scattering cross-section. This differs from single-scatter capture, where the capture rate scales as $\propto \sigma$. Furthermore, we emphasize that for a DM-baryon cross-section exceeding the geometric limit ($\sigma_{\chi N} > \sigma_{\mathrm{geo}}$, where $\sigma_{\mathrm{geo}} = 2 \times 10^{-45}~\mathrm{cm}^2$), the single-scatter description becomes invalid, rendering the associated constraints unreliable. Additionally, our expression for $\dot{P}/P$ differs from theirs.
\begin{acknowledgments}
We would like to thank Prof. Lei Wu for the useful discussions and comments. 
\end{acknowledgments}

\bibliographystyle{utphys}
\bibliography{DMfromOrbital}
 
\end{document}